\documentclass[aps,prl,reprint,longbibliography]{revtex4-1}

\usepackage{amsmath}    
\usepackage{graphicx}   
\usepackage{color}      
\usepackage{lmodern} 
\usepackage{gensymb}
\usepackage{pdfpages}
\usepackage{pgffor}

\makeatletter
\AtBeginDocument{\let\LS@rot\@undefined}
\makeatother

\begin{document}
\title{Giant electron-phonon coupling of the breathing plane oxygen phonons in the dynamic stripe phase of  La$_{1.67}$Sr$_{0.33}$NiO$_4$}

\author{A. M. Merritt$^1$, A. D. Christianson$^2$, A. Banerjee$^2$, G. D. Gu$^3$, 
A. S. Mishchenko$^{4,5}$, D. Reznik$^{1, 6}*$}

\affiliation{1. Department of Physics, University of Colorado - Boulder, Boulder, Colorado 80309, USA\\
2. Quantum Condensed Matter Division, Oak Ridge National Laboratory, Oak Ridge, Tennessee 37831, USA\\
3. Condensed Matter Physics \& Materials Science Department, Brookhaven National Laboratory, Upton, New York 11973, USA\\
4. RIKEN Center for Emergent Matter Science (CEMS), 2-1 Hirosawa, Wako, Saitama 351-0198, Japan\\
5. National Research Center ``Kurchatov Institute'', 123182 Moscow, Russia \\
6.  Center for Experiments on Quantum Materials, University of Colorado - Boulder, Boulder, Colorado, 80309, USA\\* Corresponding author: Dmitry.Reznik@colorado.edu}

\begin{abstract}
Doped antiferromagnets host a vast array of physical properties and learning how to control them is one of the biggest challenges of condensed matter physics. La$_{1.67}$Sr$_{0.33}$NiO$_4$ (LSNO) is a classic example of such a material. At low temperatures holes introduced via substitution of La by Sr segregate into lines to form boundaries between magnetically ordered domains in the form of stripes. The stripes become dynamic at high temperatures, but LSNO remains insulating presumably because an interplay between magnetic correlations and electron-phonon coupling localizes charge carriers. Magnetic degrees of freedom have been extensively investigated in this system, but phonons are almost completely unexplored. We searched for electron-phonon anomalies in LSNO by inelastic neutron scattering. Giant renormalization of plane Ni-O bond-stretching modes that modulate the volume around Ni appears on entering the dynamic charge stripe phase. Other phonons are a lot less sensitive to stripe melting. Dramatic overdamping of the breathing modes indicates that dynamic stripe phase may host small polarons. We argue that this feature sets electron-phonon coupling in nickelates apart from that in cuprates where breathing phonons are not overdamped and point out remarkable similarities with the colossal magnetoresistance (CMR) manganites.

\end{abstract}

\maketitle

\section{Introduction}
Mott insulators should become metallic when extra charge carriers are introduced by doping. However, many of them remain insulating or become very poor metals with large electrical resistivity and incoherent or diffusive transport \cite{Tokura462,Cao_2018}. This behavior is particularly common in transition metal oxides that have the potential to realize novel electronic phases with interesting and exotic properties from nontrivial topologies to superconductivity \cite{Rao89}.  Poor electrical conductivity is typically associated with charge carrier localization arising from interactions between different quasiparticles \cite{zaanen_freezing_1994}. Learning how to control these interactions is challenging, especially in the presence of strong electron-electron correlations. 

Electron-phonon coupling is often involved in localization of charge carriers in crystalline materials. For example, in the case of polaron formation, the carriers locally distort the atomic lattice and the distortions trap the carriers when the electron-phonon coupling strength is large enough \cite{Alexandrov96}. A detailed understanding of both electronic and phonon channels is necessary to accurately account for such phenomena. A lot of research focused on the former \cite{zaanen_freezing_1994,Mishchenko2000}, but the latter is poorly characterized in many interesting materials.

Time-of-flight neutron scattering instruments can map the phonon spectra over hundreds of Brillouin zones, but comprehensive analysis of these datasets is extremely difficult and time-consuming. For example, small peaks in the background can be assigned to phonons. A broad peak may arise from a superposition of two or more closely-spaced peaks. Some phonons can be overlooked since most of them have appreciable structure factor only in a few zones, etc. Recently we developed new software-based data analysis that overcomes most of these and other difficulties \cite{reznik}.

We used this software to investigate the interplay between phonon modes and carrier localization in La$_{2-x}$Sr$_{x}$NiO$_4$ (LSNO), which is isostructural with La$_{2-x}$Sr$_{x}$CuO$_4$ (LSCO), the family of cuprates in which high-temperature superconductivity was first discovered. LSNO is seen as a hole-doped antiferromagnetic Mott insulator where holes are confined within two-dimensional (2D) NiO$_2$ layers in which Ni atoms form a square lattice and O atoms bridge the nearest neighbors \cite{Zaanen01,Vojta09,Takashi04}. At low temperatures doped holes segregate into lines of charge that form antiphase domain walls between antiferromagnetic regions in the form of stripes that run along the diagonal direction with respect to the Ni-O bonds. The charge stripe period in real space equals 5.36/(2$\delta$)\AA with $\delta$ $\approx$ x \cite{Tranquada1994,Petersen18}. Here we focus on the dopant concentration x = 0.33 (La$_{1.67}$Sr$_{0.33}$NiO$_4$) but our results apply to many other doping levels and materials as discussed below.

Stripe order locks the doped holes in place but when it melts above 240K the material does not become metallic \cite{Cheong94,Ulbrich12}. The electrical resistivity continues to decrease up to about 300K and then stabilizes at a constant value up to the maximum measured temperature of 600K \cite{katsufuji_optical_1996}. It was proposed that insulating behavior is caused by polarons although their direct experimental signatures have been elusive \cite{zaanen_freezing_1994,Cheong94,katsufuji_optical_1996}.

\begin{figure}[htb!]
\includegraphics[width=\linewidth,clip]{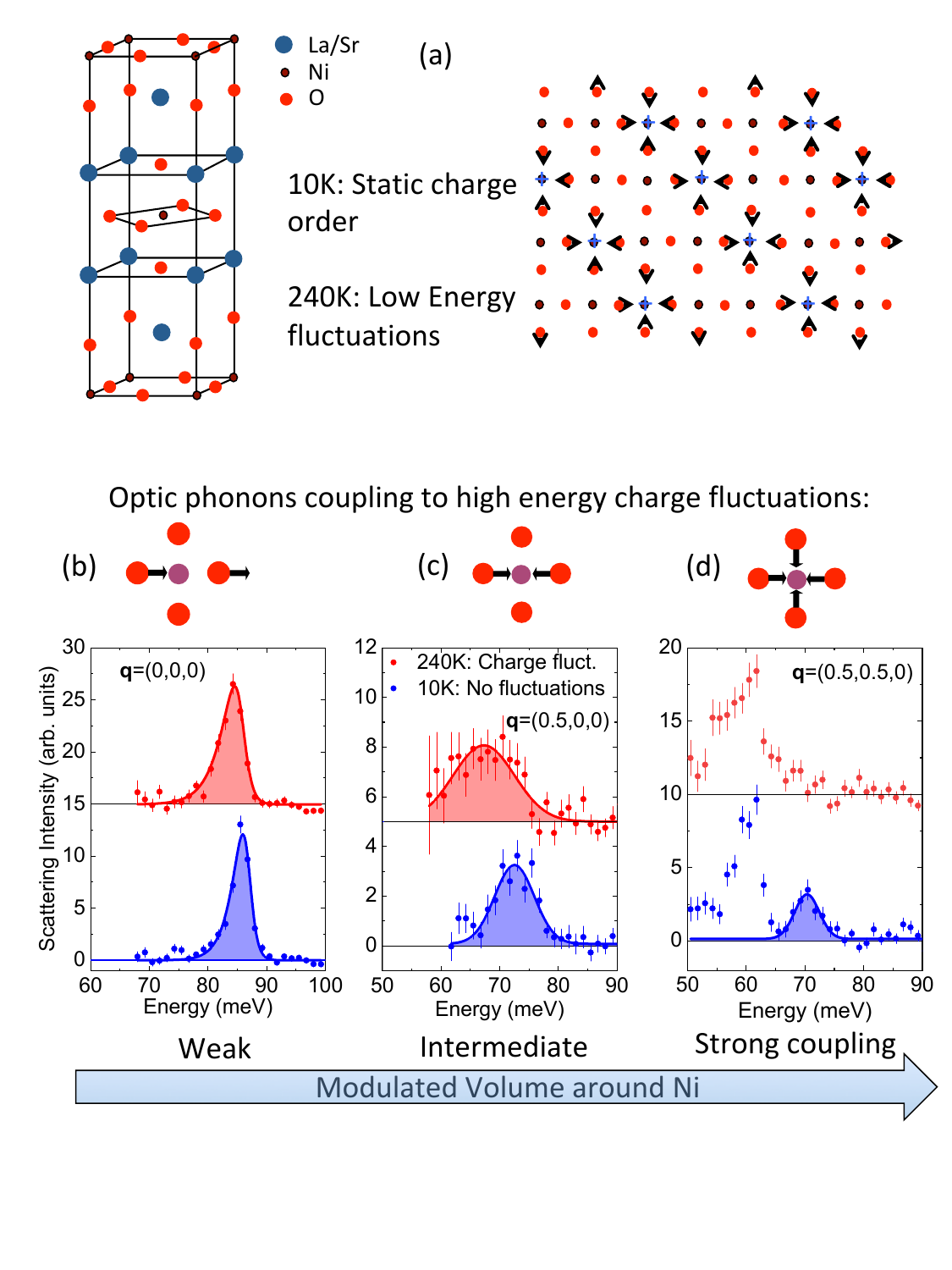}
\caption{Summary of our results. (a) Unit cell of LSNO with the schematic of the low temperature atomic lattice deformation in the NiO$_2$ planes. (b-d) Schematics represent the nickel ion surrounded by plane oxygen ions with arrows illustrating phonon eigenvectors. Highlighted peaks in the 10K and 240K spectra correspond to bond-stretching phonons of the NiO$_2$ plane: (b) Zone center where the phonon does not modulate the volume around Ni, (c) Zone boundary along [100], q=(0.5,0,0)r.l.u., (d) Zone boundary along [110], q=(0.5,0.5,0) r.l.u. Here the bond-stretching phonon can be discerned only at 10K. Peaks are asymmetric due to the shape of the energy resolution function. The arrow below (b-d) indicates the increase of modulated crystallographic volume around Ni from zero in (b) to maximum at the full breathing mode in (d).}
\label{fig:Fig1}
\end{figure}

Neutron scattering experiments revealed low energy charge fluctuations in the form of dynamic stripes illustrated in Fig.\ref{fig:Fig1}a with the largest low energy spectral weight near the charge ordering temperature \cite{Anissimova_Directobservationdynamic-2014}. Similarly to the static stripe phase, these fluctuating charges form domain boundaries between the fluctuating magnetic domains. Neutrons scatter from atomic lattice deformations, not the charges themselves, so the observed "charge" signal implies that these charge fluctuations are accompanied by dynamic atomic lattice deformations of the same wavevector, which are distinct from phonons. (See Fig. 1a)

In the present study we measured spectra of high energy phonons and found that some NiO$_2$ plane oxygen vibrations away from the Brillouin zone center are strongly damped in the dynamic stripe phase, whereas others are affected relatively weakly (Fig. \ref{fig:Fig1}c). The strongest effect is in the breathing mode, which becomes overdamped at 240K and partially recovers at higher temperatures. We argue that collapse of this mode indicates the formation of small polarons in addition to the dynamic stripes detected previously and discuss the universality of this phenomenon. 

\begin{figure}[htb!]
\includegraphics[width=\linewidth,clip]{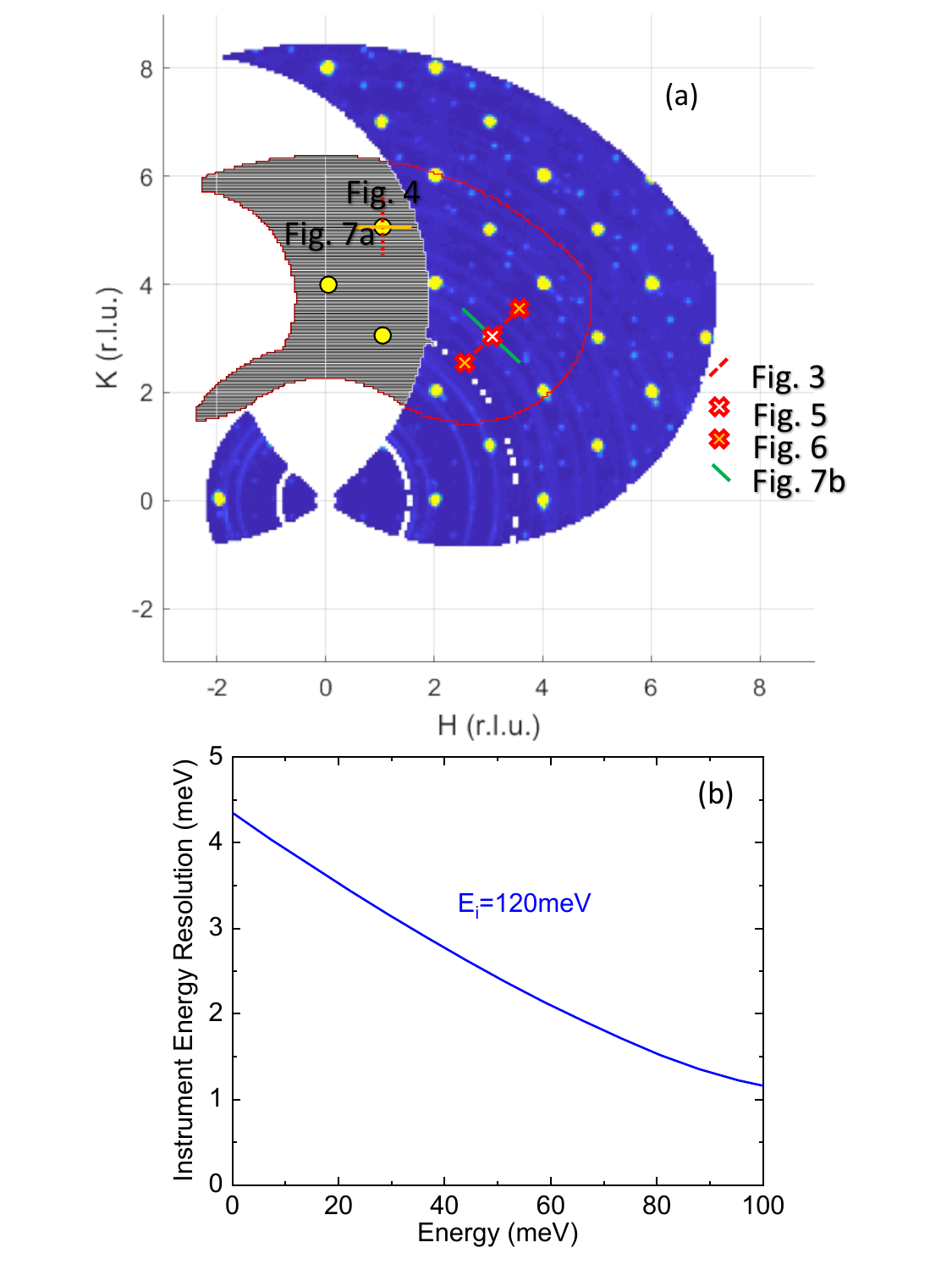}
\caption{Reciprocal space at L=0 at two energies covered in the experiment (a) and the energy resolution of the instrument (b). (a) Region with the blue background shows elastic scattering. Yellow dots originate from Bragg peaks. Area inside the red line covers the region where inelastic scattering between 88 and 92meV was measured. Lines and crosses represent the wavevectors shown in figures 3-7. (b) Full width at half maximum of the instrument resolution function.}
\label{fig:ARCS_Outline}
\end{figure}

\section{Experimental details}
Since our primary focus is on the high-temperature homogeneous phase where the stripe order is melted, we choose a unit cell containing one Ni atom per layer with the in-plane lattice parameter \textit{a} $\sim$ 3.8 \AA \ and the out-of-plane lattice  parameter \textit{c} $\sim$ 12.7 \AA \ (space group \textit{I}4/\textit{mmm}). The low temperature 3D spin ordering wavevector is then (1/2$\pm\delta$, 1/2$\pm\delta$, 0) and the charge ordering wavevector is $\bf {q}_\text{co}$ = (2$\delta$, 2$\delta$, $\pm$1) in terms of reciprocal lattice units (r.l.u.) (2$\pi$/\textit{a}, 2$\pi$/\textit{a}, 2$\pi$/\textit{c}); in our sample, $\delta$=0.33. [100] direction is defined to be parallel to the Ni-O bonds, and the [110] direction is then along the diagonal.

\begin{figure}[htb!]
\includegraphics[width=0.9\linewidth]{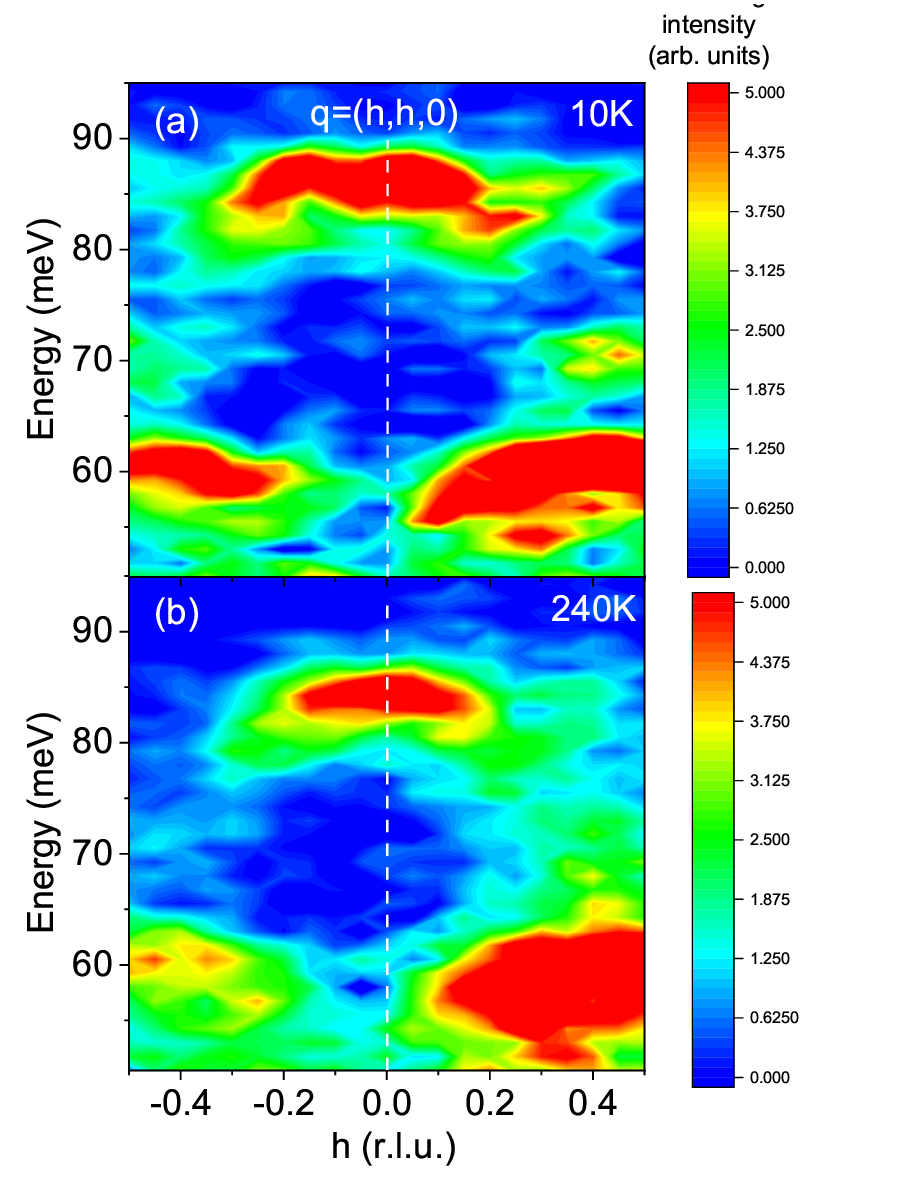}
\caption{Scattering intensity in the [110] longitudinal direction, with reduced wavevectors q=(h,h,0) at 10K (a) and 240K (b). Q=(3+h,3+h,0). The vertical dashed line denotes the zone center at Q=(3,3,0). Binning was $\pm$0.07 r.l.u. in the longitudinal direction along (h,h,0) and $\pm$0.035 r.l.u. in the transverse direction along (h,-h,0), where 1r.l.u.=2$\pi$/\textit{a}. Binning along the c-axis was $\pm$2.5 r.l.u. where 1r.l.u.=2$\pi$/\textit{c}. Note how the spectra around 70meV for $|h|>$0.3 change with temperature.
}
\label{fig:Fig2}
\end{figure}

\begin{figure}[htb!]
\includegraphics[width=0.9\linewidth]{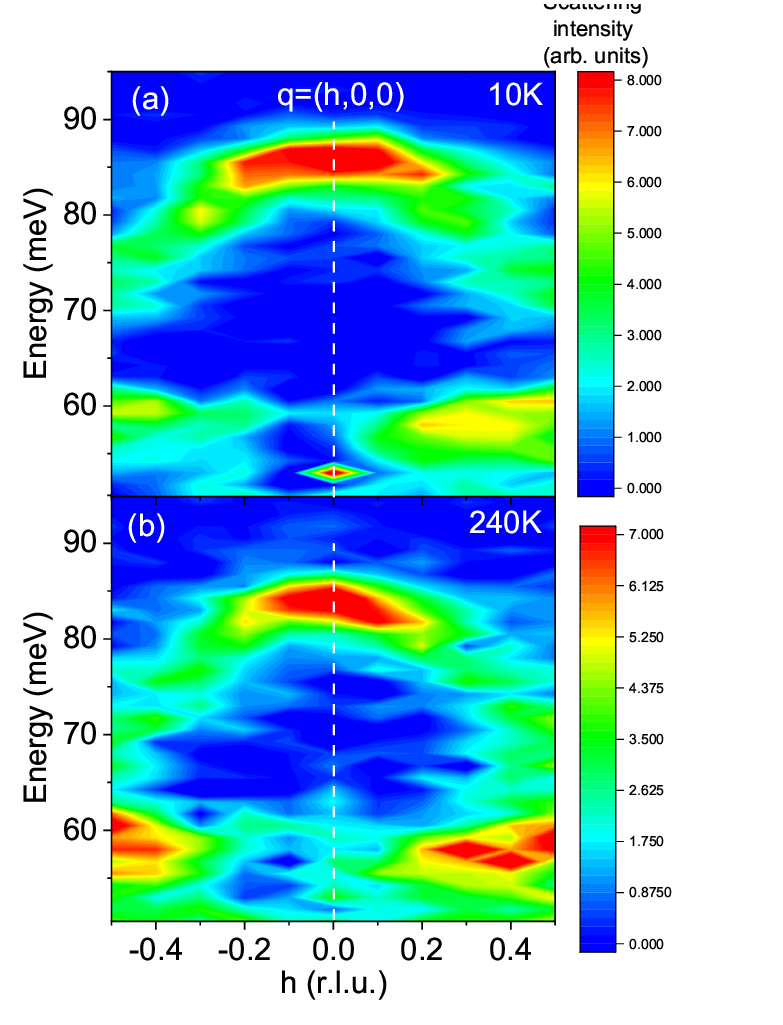}
\caption{Scattering intensity in the [100] longitudinal direction, with reduced wavevectors q=(h,0,0) at 10K (a) and 240K (b). Q=(5+h,1,0); 
Binning was $\pm$0.07 r.l.u. $\times$ $\pm$0.07 r.l.u. in the ab plane. Binning along the c-axis was $\pm$2.5 r.l.u. Note how the spectra around 73meV for $|h|>$0.3 change with temperature.
}
\label{fig:Fig3}
\end{figure}

The sample mounted with the scattering plane in the ab-plane, with the c-axis vertical and perpendicular to the beam, was measured on the ARCS spectrometer at the Spallation Neutron Source (SNS). This orientation is optimal for comprehensive measurements of phonons with atomic displacements in the Ni-O plane. Incident neutron energy E$_\text{i}$ was 120meV with the chopper running at 600Hz. T$_\text{0}$ chopper speed was 120Hz. The experiment was performed in two separate runs with the sample first mounted in the closed-cycle refrigerator set at 10K, 190K, and 240K, and in the high-temperature cryostat for the second run at 300K, 450K, and 600K. At most temperatures the sample orientation was scanned over $\sim$50$^{\circ}$ by rotating it around the vertical axis and taking neutron scattering measurements every 0.25$^{\circ}$.  The resulting dataset contained the scattering function covering over 100 Brillouin zones. (See Fig.  \ref{fig:ARCS_Outline} for coverage at L=0.) The sample rotation angular range was smaller at 450K and 600K covering reciprocal space around the 110-direction only. We used Phonon Explorer software \cite{reznik} to search for electron-phonon effects by comparing spectra at 10K with the spectra at 240K. We looked for differences between the 10K and 240K data greater than expected from conventional temperature-dependent anharmonicity. When strong effects were found, we looked at other temperatures to see if these were related to the ordering transition. The background as a function of energy was determined to be linear within uncertainty, so linear background was subtracted from the data.

The data of interest do not depend significantly on L, so we used a relatively large binning interval along L to improve statistics. The energy resolution function was broader on the low energy side than on the high energy side, hence the peaks that are close to resolution-limited are also asymmetric. 

\section{Results}

The main result of this work is summarized in Fig. \ref{fig:Fig1}: Ni-O bond-stretching phonons near the zone boundary soften and broaden (Fig. 1c) or become completely overdamped (Fig. 1d) as the stripe order melts on heating to 240K. In contrast,  phonons that do not modulate crystallographic volume around Ni such as the zone center Ni-O bond-stretching phonon (Fig. 1b) are not affected as strongly.

Figs. \ref{fig:Fig2} and \ref{fig:Fig3} show phonons at 10K and 240K dispersing in the longitudinal direction on both sides of the zone center at h=0 in the [110] and [100] directions to zone boundary  wavevectors with h=0.5. The zone centers in both figures marked with the vertical dashed lines correspond to wavevectors $\bf{Q}$=(3,3,0)/(5,1,0) in Figs. 2/3 respectively. Zone boundary wavevectors in figure 2 at Q=(2.5,2.5,0) / (3.5,3.5,0) on the left / right side of the figure both correspond to the reduced wavevector $\bf{q}$=(0.5,0.5,0). Zone boundary wavevectors in figure 3 at $\bf{Q}$=(4.5,1,0) / (5.5,1,0) on the left / right side of the figure both correspond to the reduced wavevector $\bf{q}$=(0.5,0,0). Note that the charge stripe ordering wavevector at 10K is $\bf{q}$$_{co}$=(0.33,0.33,$\pm$1)

\begin{figure}[htb!]
\includegraphics[trim={0cm 2cm 2cm 3cm},width=0.8\linewidth]{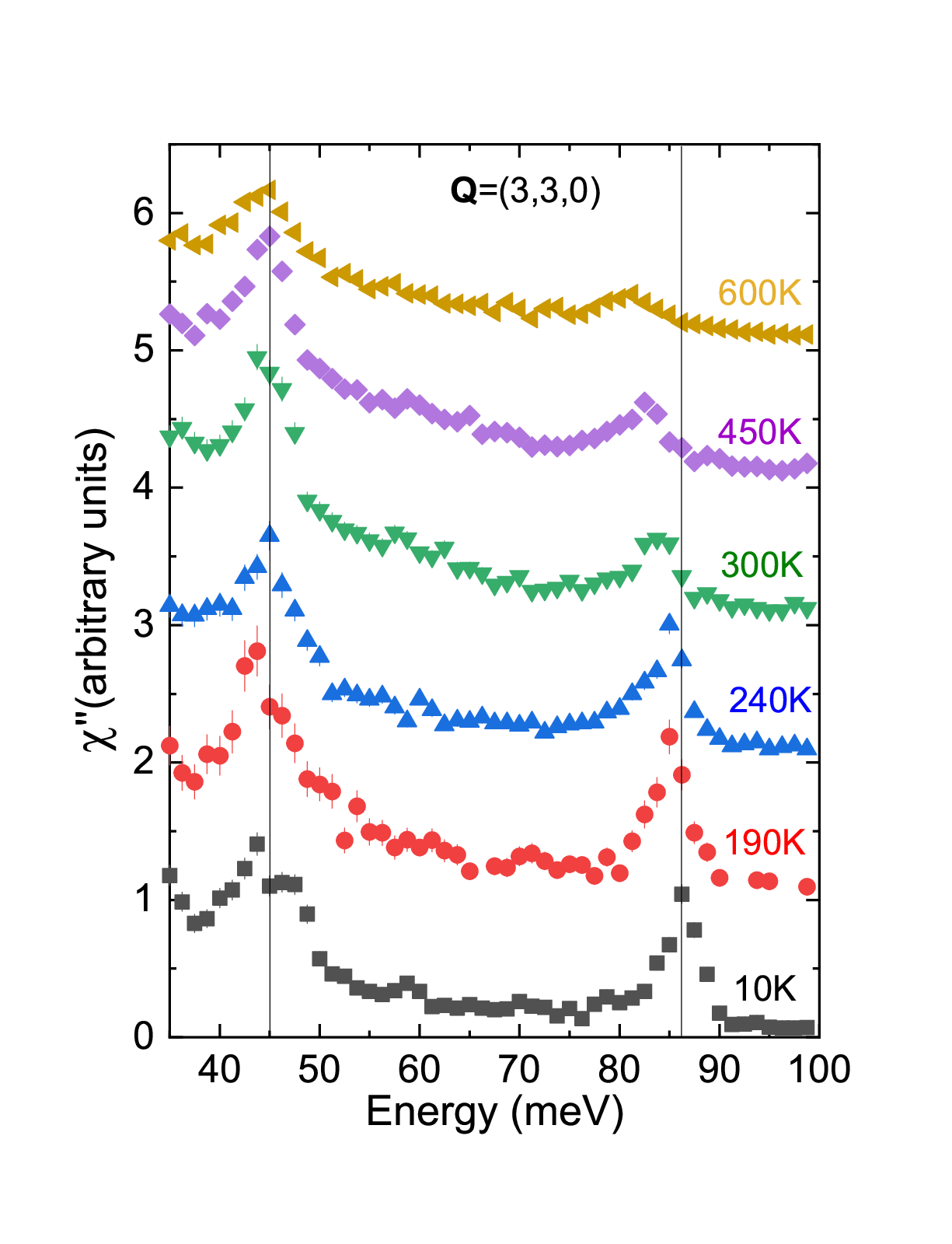}
\caption{Raw data divided by the Bose factor at the zone center wavevector Q=(3,3,0), T=240K. Binning was the same as in Fig. 3}
\label{fig:Fig5}
\end{figure}

Fig. \ref{fig:Fig5} illustrates the temperature dependence of the zone center phonons. The peaks near 45meV/85meV originate from Ni-O bond-bending/bond-stretching vibrations respectively. The 45meV zone center bond-bending phonon appears to be split into two peaks at low temperature. The splitting disappears at 190K and above, which is consistent with infrared conductivity results \cite{Homes07,Cosloviche1600735}. It was explained in terms of branch folding in the increased unit cell due to the long range stripe order. \cite{Cosloviche1600735} The bond-bending vibrations, vary relatively little with temperature up to 600K.

The IR-active bond-stretching zone center phonon at 87meV is somewhat broader than the energy resolution of the experiment (Fig. \ref{fig:ARCS_Outline}), which is consistent with the 2meV linewidth reported previously \cite{Coslovich2013}. As the atomic lattice expands on heating, the bond-stretching mode gradually shifts to lower energy and broadens due to anharmonic effects. However, it remains robust all the way up to 600K. These and other effects observed in IR reflectivity measurements occur at energy scales comparable to our energy resolution. The same applies to other phonons that we investigated, except for the bond-stretching longitudinal optic (LO) phonons away from the zone center. These have a much more pronounced temperature-dependence, which is the main focus of this paper. 

We start by discussing results at low temperature, which serve as a baseline for higher temperatures. At 10K the LO bond-stretching branch has a nearly flat dispersion in the 110 direction near the zone center, but then splits between $\bf{q}$=(0.3,0.3,0) and the zone boundary into the upper part at 85meV and the lower part at 75meV. On approach to the zone boundary, the upper part weakens, whereas the lower part intensifies. The data are consistent with earlier work \cite{tranquada_bond-stretching-phonon_2002}.

The LO branch disperses downwards and broadens towards the zone boundary in the [100] direction as previously reported \cite{tranquada_bond-stretching-phonon_2002}. Our measurements suggest that this branch also splits into two (Fig. \ref{fig:Fig3}).  Although these cannot be fully resolved in a single scan, we observed the variation of the lineshape from zone to zone, which points at two closely-spaced branches with different eigenvectors. 

The lowest branch in Figs 2,3 around 60meV originates from apical oxygen vibrations along the \textit{c}-axis \cite{Stoichiometric/NonStoichiometric_2001}. It mixes with the Ni-O bond-bending vibrations away from the zone center, which is responsible for nonzero scattering intensity observed in our measurements. This branch does not show strong coupling to charge fluctuations and we will not discuss it further.

\begin{figure}[htb!]
\includegraphics[trim={1.5cm 4cm 1cm 3cm},width=\linewidth]{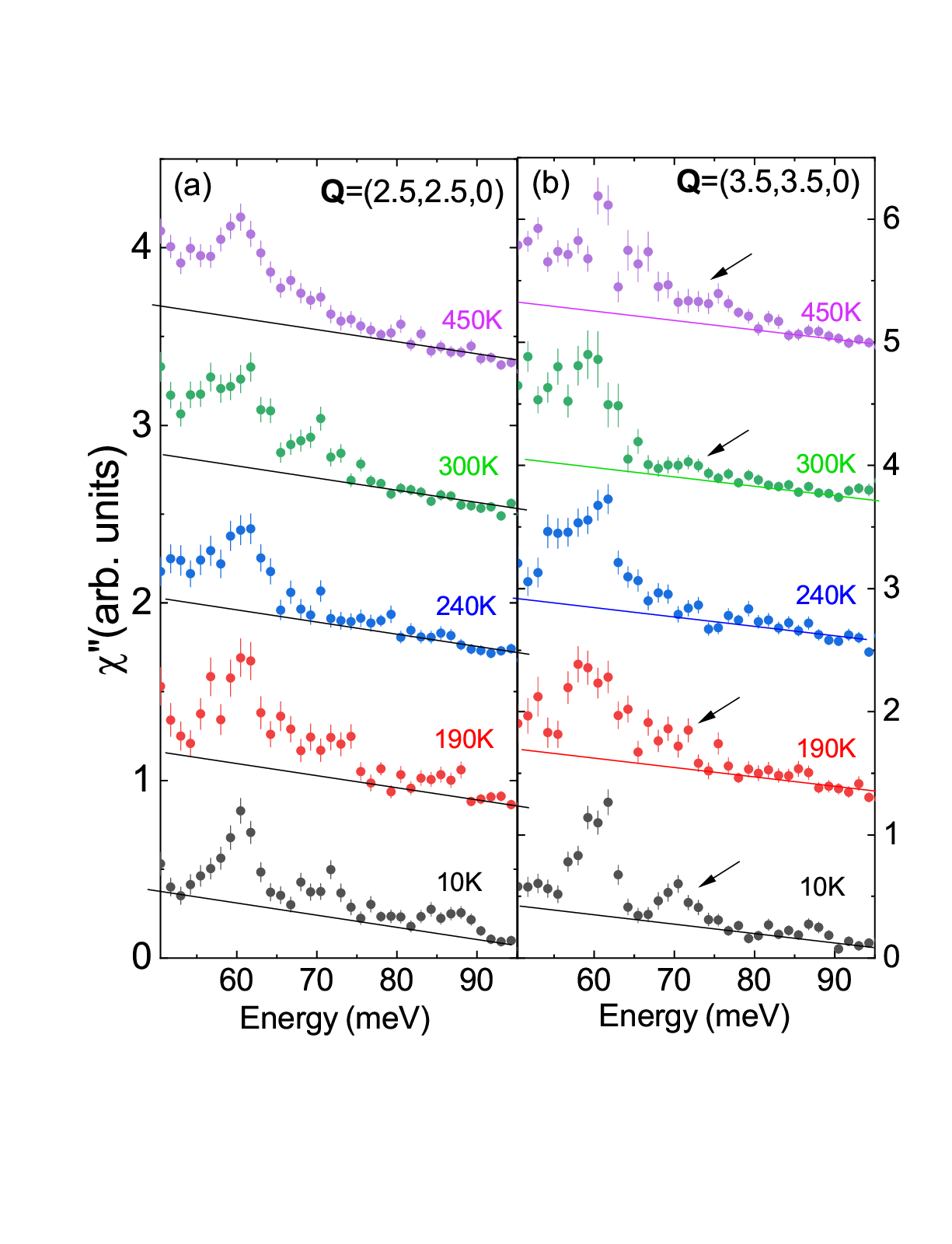}
\caption{Raw data divided by the Bose factor at two zone boundary wavevectors: Q=(2.5,2.5,0) (a) and Q=(3.5,3.5,0) (b). The two wavevectors correspond to the same reduced wavevector q=(0.5,0.5,0). Straight lines represent the background. Binning was the same as in Fig. 2. Arrows point at the peak from the breathing phonon that is well defined at low temperature, becomes overdamped at 240K, and reappears at higher temperature. The same behavior is observed in (a) and (b).}
\label{fig:Fig4}
\end{figure}

Our primary focus is on the temperatures of 240K and above where we observe a single strongly renormalized branch in all directions. The LO Ni-O breathing bond stretching branches away from the zone center broaden and soften half-way to the zone boundary on heating and branch splitting is no longer observed. The phonon softening from 10K to 240K increases from 1meV at the zone center to about 4meV at the zone boundary [$\bf{q}$=(0.5,0,0)] and the peak broadens substantially (Fig. \ref{fig:Fig3}) Fig. \ref{fig:Fig1}c shows the same effect in another Brillouin zone at $\bf{Q}$=(5.5,0,0), which rules out any spurious origin of the anomaly. On approach to the zone boundary M point [$\bf{q}$=(0.5,0.5,0)] in Fig. \ref{fig:Fig2} phonon renormalization is even larger: The bond-stretching phonon evolves from a relatively narrow profile at 10K peaked at 74meV to a broad barely discernible overdamped lineshape at 240K (Also see Fig. \ref{fig:Fig1}d). 

Figure \ref{fig:Fig4} shows the full temperature-dependence of the phonon spectrum at ${\bf q}$=(0.5,0.5,0) up to 450K in two different Brillouin zones. The phonon at 74meV broadens at 190K where magnetic order melts and is completely washed out at 240K where charge order disappears. It recovers partially at 300K and 450K.

\begin{figure}[htb!]
\includegraphics[trim={0cm 1cm 8cm 0cm},width=1.0\linewidth]{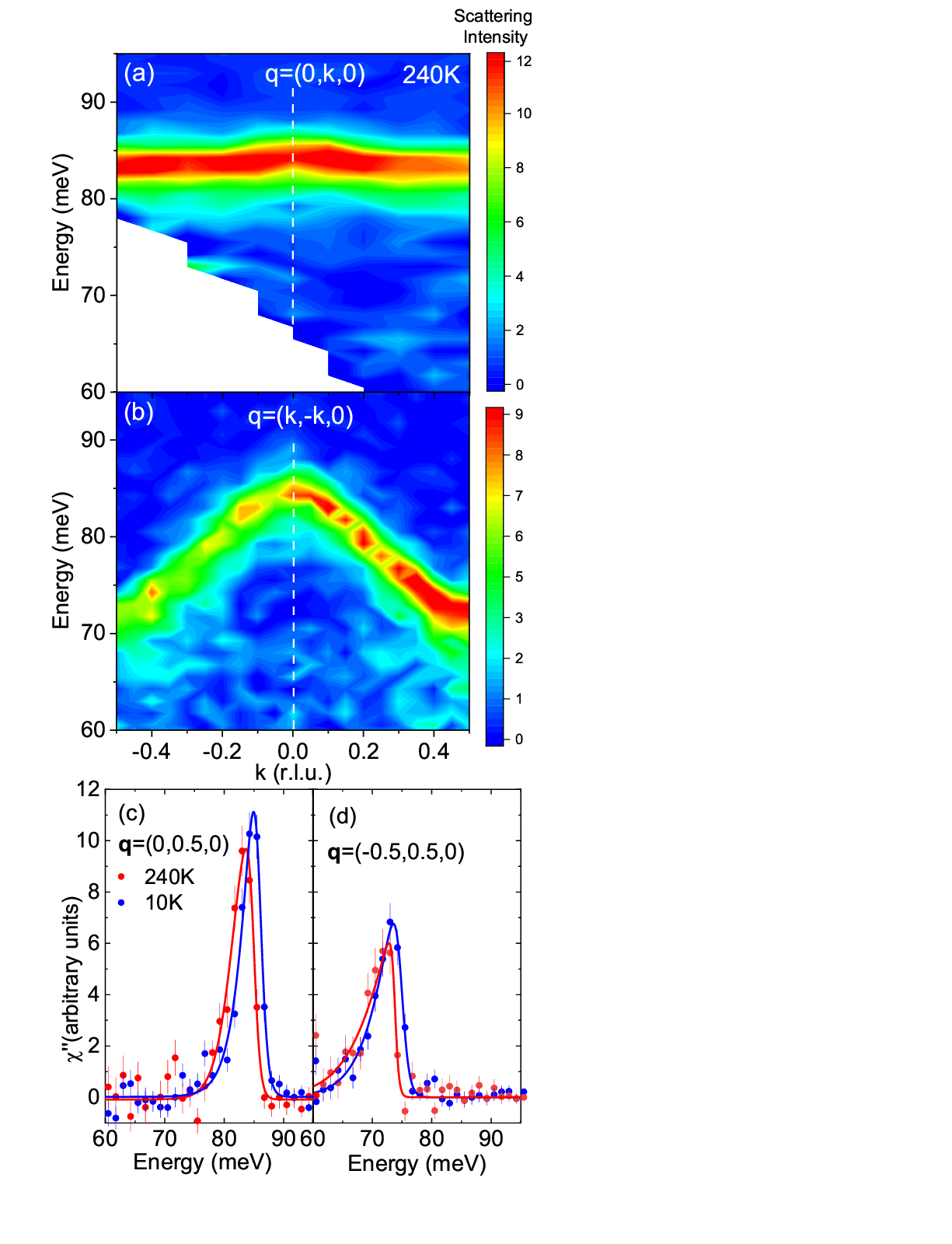}
\caption{Phonon dispersion in the transverse directions. (a) along the Ni-O bond at wavevectors Q=(5,1+k,0) (Binning was the same as in Fig. 3); (b) along the diagonal direction, ${\bf Q}$=(3-k,3+k,0) (Binning was the same as in Fig. 2); (c) Zone boundary phonons from (a) at 10K and 240K (d) Zone boundary phonons from (b) at 10K and 240K. Lines in c,d are guides to the eye }
\label{fig:Fig7}
\end{figure}

This behavior can be understood if one considers the volume of the oxygen octahedron around Ni. For the zone center phonon, the deformation due to lattice vibrations does not change the volume around Ni. The modulated volume increases precisely as a sine function between the zone center and the zone boundary. Fig. \ref{fig:Fig7} confirms the connection between the phonon anomaly and the modulation of the octahedron volume around Ni: The transverse branches involving stretching of the same Ni-O bonds that do not modulate the volume around Ni at any wavevector are narrow throughout the Brillouin zone at 240K, not just at the zone center. 


\section{Discussion}

LSNO received a lot of attention as the analogue of the high T$_c$ cuprates. Undoped cuprates are insulating antiferromagnets that become metallic and superconducting when doped. They are also characterized by a low temperature charge density wave CDW \cite{Tranquada1995,Ghiringhelli821,Blackburn13}, but the associated atomic lattice distortions are much smaller than in LSNO. Futhermore, long range magnetic order is established only in a few special cases, presumably because cuprates are spin-1/2 systems with large fluctuations.  LSNO at low temperatures is an insulating spin 1 antiferromagnet with small magnetic fluctuations and a large atomic lattice deformation that aides in trapping the doped holes in magnetic domain boundaries. These properties make the low temperature phase of LSNO closer to updoped cuprates that to the metallic ones. 

In the high temperature phase of LSNO, charge carriers are not gapped and optical conductivity has a pronounced midinfrared absorption band similar to doped cuprates. This phase is also characterized by pronounced low energy charge fluctuations in the form of stripes. They have many similarities to the dynamic component of the CDW in cuprates, which is strong and possibly important for the pseudogap and superconductivity \cite{miao_incommensurate_2018}. Thus the high temperature phase of LSNO is the one to study as an analogue of the cuprates. Our work highlights and characterizes strong electron-phonon interaction in this phase. 

The electron-phonon interaction can be described by the following Hamiltonian:
\begin{equation}
H_{int} = \frac{1}{\sqrt{N}} \sum_{\bf q k} \Gamma({\bf q,k}) 
(c^{\dagger}_{\bf k} c_{\bf k-q} b_{\bf q} + c^{\dagger}_{\bf k-q} c_{\bf k} b^{\dagger}_{\bf q}) \; . 
\label{Ham}
\end{equation}
It represents a process where an electron changes its momentum by ${\bf q}$ when it annihilates/creates ($b_{\bf q}$/$b^{\dagger}_{\bf q}$ a phonon with momentum ${\bf q}$. $c^{\dagger}_{\bf k} / c_{\bf k}$ are creation/annihilation operators of electron with momentum ${\bf k}$, $\Gamma$  is the electron-phonon coupling.

Apart from special cases, the interaction $\Gamma({\bf k,q}) \approx \Gamma({\bf q})$
depends only on ${\bf q}$. In the case of long-range Fr\"ochlich Hamiltonian arising from the Coulomb interaction, the strongest coupling is at ${\bf q}$=0:

\begin{equation}
\Gamma({\bf q}) \sim \frac{\sqrt{\alpha}}{q^{(d-1)/2}} \;
\label{Froh}
\end{equation}
where $\alpha$ is a dimensionless coupling constant and $d =3$ or $d=2$ is the dimension of the system 
\cite{Peeters_et_al_1986}. This interaction should have the biggest effect on phonons near the zone center, which is inconsistent with our findings.

As an alternative, Holstein type coupling based on short-range electron-phonon interaction is characterized by momentum independent $\Gamma$: \cite{Holstein_1959}
 
\begin{equation}
\Gamma({\bf q}) = g
\label{Holst}
\end{equation}
 
However it is also inconsistent with our results, which show strong $\bf{q}$-dependence of electron-phonon coupling. 

Instead our results point at a breathing-type electron-lattice interaction peaked at the zone boundary \cite{Kikoin_1990} and characterized by the coupling of electronic charge fluctuations to vibrations of lighter ions modulating the volume around a heavier one. It is similar to coupling to nonzero wavevector charge fluctuations discussed in \cite{Allen_1977,Bilz_et_al_1979}. Strong manifestations of this type of interaction were first experimentally observed \cite{Mook78,Mook82}
and explained \cite{Alekseev_1989,Kikoin_1990} in mixed valence compounds. In our case 
\begin{equation}
\Gamma({\bf q}) \sim g \sin \frac{q}{2}
\label{Breath}
\end{equation}
which peaks at the Brillouin zone boundary \cite{Kikoin_1990,Slezak_et_al_2006}.

The important difference between our findings and models of Ref. \cite{Allen_1977,Bilz_et_al_1979} is the observation of strong broadening of the phonons not accounted for in purely harmonic models based on dynamical matrices. The adiabatic part of the electron-lattice coupling takes into account high-energy electronic excitations leading to well defined phonon modes of zero width. It was noticed long ago \cite{BroKag} and clearly re-established recently \cite{GiustinoReview} that the existence of low-lying electronic excitations, whose energy is comparable with that of the lattice vibrations is necessary for phonon broadening caused by electron-phonon interaction. In this case, the nonadiabatic part of the interaction caused by soft electronic excitations leads to the damping of otherwise perfectly stable lattice vibrations. 

Thus our results point at an appearance of soft electronic excitations at energies comparable to the Debye energy in the dynamic stripe phase, which is consistent with optical measurements \cite{katsufuji_optical_1996}. These charge excitations extend to low energies where they are pinned by the dynamic stripe domain boundaries \cite{Anissimova_Directobservationdynamic-2014}. 

At higher energies magnetic fluctuations are featureless in ${\bf q}$-space and charge fluctuations can interact with phonons.  A relatively small broadening of about 2meV as well as the Fano lineshape with unusual nonequilibrium dynamics has been reported for the zone center bond-stretching phonons, which have zero breathing character, based on infrared reflectivity measurements \cite{Coslovich2013}. We showed that the interaction strength increases dramatically with the breathing character of the phonons away from the zone center. These observations point at a complex highly cooperative behavior between charge, spin, and lattice degrees of freedom.

Recent RIXS experiments showed considerable electronic charge character in cuprates associated with an analogous Cu-O bond stretching phonon \cite{chaix_dispersive_2017}, which softens with doping on approach to the zone boundary \cite{PINTSCHOVIUS1991156,Pintschovius_review_2005,Stoichiometric/NonStoichiometric_2001}. In both cuprates and nickelates, holes reside primarily on the oxygen orbitals \cite{kuiper_unoccupied_1991}. However, in the nickelates the Ni character of the doped holes \cite{Merritt19}, which tends to attract the surrounding O ions, is stronger than the Cu character in cuprates \cite{fabbris_doping_2017}. This difference enhances the breathing character of electron-phonon coupling in the nickelates.

We now show that strong electron-phonon coupling of the breathing phonons favors small polaron formation in the dynamic stripe phase of LSNO. 

Long-range electron-phonon interaction represented by the Fr\"ochlich Hamiltonian leads to large polarons where a single electron or hole distorts the surrounding atomic lattice spanning many unit cells. Carrier-carrier interactions prevent the formation of such polarons when the carrier density is high. 

A number of theoretical models propose small polarons, which can form at high carrier densities due to their small size. They arise from short range carrier-lattice interactions and involve only a few unit cells or even just one unit cell \cite{Holstein_1959,Emin93}. For all types of polarons most of the electronic quasiparticle spectral weight is pushed below the Fermi surface and electrical conductivity is suppressed.

One may think that the Holstein Hamiltonian \cite{Holstein_1959} is most favorable to small polarons since it is already purely local: It is nonzero at the position of the electron and zero on neighboring sites \cite{Devreese_2009}. In fact it was shown theoretically that as the interaction strength increases, the Holstein interaction has a much sharper transition into the strong-coupling regime necessary for polaron formation than the long range Fr\"ochlich interaction \cite{Berciu}. However, the breathing interaction involves volume contraction on a site accompanied by an anti-phase volume expansion on neighboring sites, i.e. the electron is attracted to the site it occupies while being repelled from surrounding sites. This interaction has a stronger on-site confinement than the Holstein interaction, which is zero on the neighbors. Indeed the transition into the strong-coupling regime with increasing coupling strength is even sharper for the breathing type coupling \cite{Berciu} implying that the formation of small polarons is most probable in the case of the breathing interaction. 

Slowly fluctuating dynamic stripes are essentially static at the energy scale of midinfrared conductivity (order of 0.5eV) and play a role of an impurity potential. The optical response of such trapped carriers is significant only at high energies \cite{de_Candia_2019}. However, midinfrared optical conductivity in the high-temperature dynamic stripe phase is characterized by a broad absorption band extending all the way to zero energy. At low energies optical conductivity can be fit by the small polaron model \cite{katsufuji_optical_1996}, which is consistent with coexistence of small polarons with dynamic stripes. 

Based on our observations and theoretical considerations, we conjecture that small polarons form in the high temperature phase of LSNO. To prove it, angle resolved photoemission measurements (ARPES) in the high temperature phase should show that most of the electronic quasiparticle spectral weight is pushed below the Fermi level \cite{Sun11799} and has sidebands that correspond to the breathing phonons.

Special role of the phonons that modulate the volume around the metal ions has also been reported in a colossal magnetoresistance (CMR) manganese oxide, La$_{1-x}$Ca$_{x}$MnO$_3$ (LCMO) \cite{Jhang_Manganite2001}.

CMR manganites are metallic ferromagnets at low temperatures. Their electrical resistivity jumps dramatically above the Curie temperature characterized by short range charge/orbital order often described in terns of large Jan-Teller polarons. \cite{Dai2000} These collective polarons are fundamentally different from large or small polarons discussed above. Here electron-phonon and electron-electron interactions underlie short-range-ordered states that localize charge carriers that are sometimes called polaronic \cite{Tokura462}. Such polarons are characterized by specific nonzero wavevectors, $\bf{q}$, and extend over many unit cells. In this context a polaron necessarily involves many electrons interacting with each other and the atomic lattice. Localization occurs due to quasiperiodic cooperative modulations of electronic charge density locked to a deformation of the atomic lattice of the same wavevector. In manganites such polarons are observed by neutron scattering as an elastic or quasielastic broad peak centered at the transverse wave vector $\bf{q}$=(-1/4,1/4,0) \cite{PhysRevLett.85.3954}. 

Polarons in CMR manganites are analogous to dynamic charge stripes in the nickelates. Both appear in the high temperature phases at low energies, are centered at specific, although different, wavevectors, and involve atomic lattice deformations induced by charge inhomogeneity. An important difference is that magnetic domains underlie stripe formation in nickelates, but charge/orbital order in the manganites does not directly involve magnetic degrees of freedom.

Ref. \cite{Jhang_Manganite2001} shows that LO bond-stretching modes become overdamped in the high temperature phase of LCMO similarly to the behavior we report here for LSNO. We observed this phonon effect much more clearly in another manganite, La$_{1-x}$Sr$_{x}$MnO$_3$ (LSMO) where LO Mn-O bond-stretching phonon becomes overdamped on approach to the zone boundary (see Supplementary Material). Striking similarity in the breathing phonons between nickelates and manganites with very different low energy physics indicates that small polarons based on breathing LO modes may be generic in doped perovskite oxides and possibly other systems with similar structure such as the photovoltaic perovskites. 

In most copper oxide superconductors bond-stretching phonon branches exhibit strong electron-phonon anomalies near $\bf{q}$=(1/4,0,0) \cite{PintschoviusOverdoped,Reznik08,REZNIK201275,Pintschovius_review_2005}. In the copper oxide family based on La$_{2-x}$Sr$_{x}$CuO$_4$, which is isostructural to LSNO, bond-stretching phonons broaden and soften near this wavevector with subtle differences between compounds with and without static stripe order \cite{Reznik2007,REZNIK20083103}. Zone boundary LO phonons in the copper oxides are not overdamped \cite{PintschoviusOverdoped}, which indicates that the electron-phonon interaction of the breathing type is much weaker than in the nickelates and manganites so that small polarons do not form \cite{bi_polaron_1993}. It is intriguing that there is no apparent phonon anomaly associated with $\bf{q}$$_{co}$ in the prototypical and very robust charge stripe phases in LSNO, but it can be seen for the much weaker charge stripe phase in the cuprates. This observation indicates that stripe phases in the two compounds may be fundamentally different and requires further investigation.

We demonstrated that electron-phonon coupling is very strong for breathing Ni-O modes in the high temperature phase of LSNO. This interaction favors small polarons of the breathing type that, if they form, would coexist with dynamic stripes and make the system more susceptible to correlations and localization. We hope our work will stimulate the inclusion of this interaction into theoretical models of materials with correlated electrons. Tuning it opens an additional way to control their electronic properties.

A.M.M. and D.R. would like to thank J.M. Tranquada for helpful discussions. This work used Phonon Explorer software for data analysis. Work at the University of Colorado-Boulder was supported by the U.S. Department of Energy, Office of Basic Energy Sciences, Office of Science, under Contract No. DE-SC0006939. A.S.M. was supported by JST CREST Grant Number JPMJCR1874, Japan. Work at Brookhaven National Laboratory was supported by the U.S. Department of Energy, Office of Basic Energy Sciences, under Contract No. DE-SC0012704. This research used resources at the Spallation Neutron Source, a DOE Office of Science User Facility operated by the Oak Ridge National Laboratory.

\bibliography{bibli}{}

\end{document}